\documentstyle[prl,aps,floats,epsf]{revtex}

\begin{document} 
\draft

\twocolumn[\hsize\textwidth\columnwidth\hsize\csname
@twocolumnfalse\endcsname

\preprint{HEP/123-qed}

\title{Finite Temperature Effects in One-dimensional Mott-Hubbard 
Insulator: Angle-Resolved Photoemission Study of 
Na$_{0.96}$V$_{2}$O$_{5}$}

\author{K. Kobayashi, T. Mizokawa, A. Fujimori} 

\address{Department of Physics, University of Tokyo, Hongo 7-3-1, 
Bunkyo-ku, Tokyo, 113-0033, Japan}

\author{M. Isobe, Y. Ueda} 

\address{Materials Design and Characterization Laboratory, Institute 
for Solid State Physics, University of Tokyo, Roppongi 7-22-1, 
Minato-ku, Tokyo, 106-8666, Japan}

\author{T. Tohyama and S. Maekawa} 

\address{Institute for Materials Research, Tohoku University, Katahira 
2-1-1, Aoba-ku, Sendai, 980-8577, Japan}

\date{Phys. Rev. Lett., 82, 803 (1999)}
\maketitle

\begin{abstract}
We have made an angle-resolved photoemission study of a 
one-dimensional (1D) Mott-Hubbard insulator Na$_{0.96}$V$_{2}$O$_{5}$ 
and found that the spectra of the V $3d$ lower Hubbard band are 
strongly dependent on the temperature.  We have calculated the 
one-particle spectral function of the one-dimensional $t$-$J$ model at 
finite temperatures by exact diagonalization and compared them with 
the experimental results.  Good overall agreement is obtained between 
experiment and theory.  The strong finite temperature effects are 
discussed in terms of the existence of the ``Fermi surface'' of the 
spinon band.
\end{abstract}

\vskip1pc]

\narrowtext


The most striking and non-trivial theoretical prediction for 1D 
strongly correlated systems is the spin-charge 
separation~\cite{LiebPRL68}: the degrees of freedom of an electron are 
decoupled into elementary excitations of spin and charge called 
``spinon'' and ``holon'', respectively.  Angle-resolved photoemission 
spectroscopy (ARPES) is a powerful technique to study the spin-charge 
separation: Kim \textit{et al.}~\cite{KimPRL96} performed a pioneering 
ARPES work on the 1D charge-transfer insulator SrCuO$_{2}$ and 
found that the spectra agree well with the theoretical one-particle 
spectra of the 1D $t$-$J$ model with realistic parameters, identifying 
such spinon and holon excitations.  We subsequently investigated 
whether the same scenario is applicable for a 1D Mott-Hubbard type 
insulator NaV$_2$O$_5$~\cite{KobayashiPRL98}.

Strongly-correlated systems are often characterized by the presence of 
a characteristic low energy scale in spite of the large energy scales 
of bare interaction strengths.  For the Hubbard model, the relevant 
low energy scale is set by the superexchange interaction $J \sim 
4t^{2}/U$ ($\ll t, U$) rather than the bare interaction parameters, 
the transfer integral $t$ and the on-site Coulomb repulsion $U$.  
Since the photoemission spectrum is a projection of the initial state 
onto the set of final states, drastic finite-temperature effects may 
be expected for a temperature change of the order of such a 
characteristic low energy scale.  Finite-temperature effects can be 
particularly drastic in 1D systems~\cite{PencPRB97} because of the 
existence of the ``Fermi surface'' of spinon excitations.  In this 
Letter, we present the result of a temperature-dependent ARPES study 
on NaV$_2$O$_5$.  While severe charging effects had previously 
prevented the measurements of NaV$_2$O$_5$ below $\sim 300$ 
K~\cite{KobayashiPRL98}, more conductive 
Na$_{0.96}$V$_2$O$_5$~\cite{IsobeJAC97} enabled us to obtain ARPES 
spectra at temperatures as low as $\sim 120$ K. Dispersive features of 
V 3\textit{d} character in the lower Hubbard band were found to be 
dramatically dependent on the temperature.  In addition, we have made 
a comparison with theory and confirmed that the observed 
finite-temperature effect is due to strong correlation effect rather 
than the simple thermal broadening.

Recently NaV$_2$O$_5$ has widely aroused much interest as a quasi-1D 
quantum-spin system.  Crystalographically, it is still controversial 
whether the V atoms in VO$_{5}$ pyramids running along the 
\textit{b}-axis of NaV$_2$O$_5$ is mixed-valent 
($\textrm{V}^{4+}:\textrm{V}^{5+}=1:1$)~\cite{CarpyActaCryst77,IsobeJPSJ96-1} 
or uniform-valent (V$^{4.5+}$)~\cite{SmolinskiPRL1998}.  In the former 
case, this compound can be no doubt regarded as a half-filled 
chain~\cite{IsobeJPSJ96-1}, while in the latter case it is viewed as a 
quarter-filled ladder system~\cite{SmolinskiPRL1998,OhamaPreprint}.  
At low temperatures around or below its spin-Peierls-like (SP) 
transition temperature $T_{SP} \sim 34$ K~\cite{IsobeJPSJ96-1}, the 
difference between those two models is significant in terms of the 
charge ordering pattern.  On the other hand, the magnetic 
susceptibility $\chi(T)$ well above $T_{SP}$ is successfully fitted to 
the Bonner-Fischer curve with $J \sim 560$ K~\cite{IsobeJPSJ96-1}, 
indicating that NaV$_2$O$_5$ behaves as a good 1D antiferromagnetic 
Heisenberg chain in this temperature region.  In fact, it is 
theoretically supported that it can be mapped on to the 1D Heisenberg 
chain~\cite{SmolinskiPRL1998}.  Except that the SP transition is 
suppressed~\cite{IsobeJAC97}, the Na-deficient Na$_{0.96}$V$_2$O$_5$ 
has almost the same magnetic properties as 
NaV$_2$O$_5$~\cite{CommentOnCorrelationLength}.  Though remaining an 
insulator, Na$_{0.96}$V$_2$O$_5$ is more conductive than NaV$_2$O$_5$ 
due to the doped holes.

Single crystals of Na$_{0.96}$V$_{2}$O$_{5}$ were prepared as reported 
in Ref.~\cite{IsobeJCG97}.  They could be easily cleaved parallel to 
the \textit{ab} plane.  The ARPES measurements were made using the He 
I resonance line ($h\nu = 21.2$ eV) and a hemi-spherical analyzer with 
angular resolution $\pm 1^{\circ}$ and energy resolution $80$-$100$ 
meV. The measurement temperature ranged from $T=120$ K ($= 0.21J$) to 
room temperature 300 K ($= 0.54J$).  The measurements were performed 
for several \textit{in situ} cleaves, for which we carefully cycled 
the temperatures of cleavage and measurements in order to exclude 
any extrinsic effects such as surface degradation and contamination.

Before discussing the V 3\textit{d} band features of our main 
interest, we would like to mention that the O 2\textit{p} band 
structure was found very anisotropic: ARPES spectra with momentum 
parallel to the \textit{b}-axis ($k \parallel b$) show rich dispersive 
features while those with momentum parallel to the \textit{a}-axis ($k 
\perp b$) have no dispersions, supporting the one-dimensionality of 
this compound.  These results well agree with those of stoichiometric 
NaV$_{2}$O$_{5}$~\cite{KobayashiPRL98}, indicating that the Na 
deficiency has no appreciable influence on the O 2\textit{p} band 
structure.  Furthermore, the O 2\textit{p} spectra have no obvious 
temperature-dependent changes, making a clear contrast to the 
remarkable changes of the V 3\textit{d} band described below.

\begin{figure} [ht]
\center \epsfxsize=82mm \epsfbox{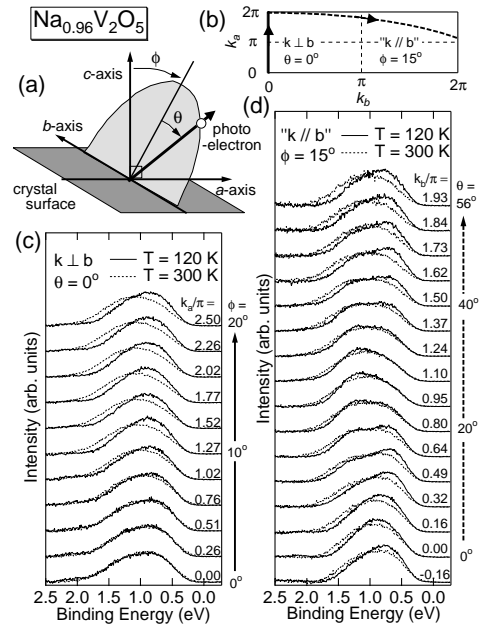} 
\vspace{1pt} 
\caption{(a) Definition of the take-off angles $\phi$ and $\theta$ of 
photoelectrons.  (b) Measured cuts in the the Brillouin zone.  (c) 
Spectra for the $k \perp b$ cut and (d) those for the ``$k \parallel 
b$'' cut.  Solid and dashed curves show spectra taken at 120 K and 300 
K, respectively.  Note that each spectrum is normalized to its area 
and that the absolute intensities of the spectra near $\theta = \phi = 
0^{\circ}$ are much weaker than the others.}
\label{ExpResult}
\end{figure}

Figures~\ref{ExpResult} (a) and (b) show the geometry of the 
measurements.  By scanning the momentum in the Brillouin zone (BZ) as 
indicated, we obtained ARPES spectra along the $k \perp b$ and ``$k 
\parallel b$'' cuts as shown in Figs.~\ref{ExpResult} (c) and (d), 
respectively.  Hereafter, $k_{a}$ and $k_{b}$ are the momenta along 
the \textit{a}- and \textit{b}- axes in units of the inverse of the 
corresponding lattice constant, respectively.  All the spectra have 
been normalized to the area of the lower Hubbard band of V 3\textit{d} 
origin between the binding energies ($E_{B}$) of 0.0 and 2.5 eV after 
subtracting the integral background.  We notice that the $k \perp b$ 
cut [Fig.~\ref{ExpResult}(c)] shows a clear temperature dependence and 
a weak angle dependence except for around the $\Gamma$ point 
($\theta=\phi=0^{\circ}$).  In addition, the absolute intensity at the 
$\Gamma$ point was found to be weaker than half of those at $\phi > 
10^{\circ}$.  These observations can be explained by the facts that 
the occupied V 3\textit{d} orbital has $xy$ symmetry lying 
approximately in the $ab$-plane~\cite{OhamaJPSJ97} and therefore that 
normal emission from this orbital is forbidden due to selection 
rules~\cite{HufnerPS94}.  We therefore conclude that the temperature 
dependence results from the intrinsic finite temperature effects of 
the V 3$d_{xy}$ band~\cite{CommentOnResidual} and that the momentum 
dependence along the $k \perp b$ direction is due to the matrix 
element effects of the $d_{xy}$ orbital.

\begin{figure} [ht]
\center \epsfxsize=85mm \epsfbox{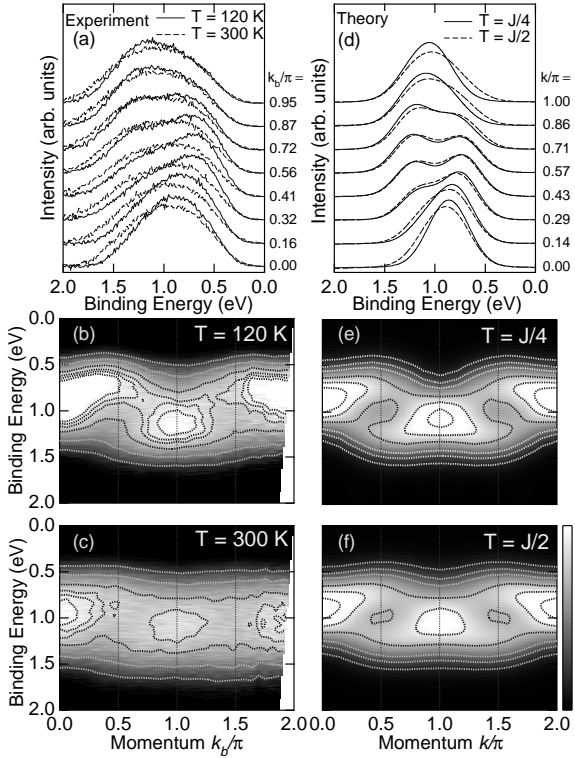} 
\vspace{1pt} 
\caption{Comparison between the experimental and theoretical spectra.  
Panel (a) shows the ARPES spectra of Na$_{0.96}$V$_{2}$O$_{5}$ with 
the intensity plots in panels (b) and (c).  Panels (d), (e) and (f) 
show the theoretical counterparts calculated for the half-filled 
14-site $t$-$J$ model.  Here $J = 0.05$ eV.}
\label{CompExpTheory}
\end{figure}

Figure~\ref{ExpResult} (d) shows the result for the ``$k \parallel 
b$'' cut.  In order to avoid the $\Gamma$ point where the 
matrix-element effects prohibit emission from $d_{xy}$, the $k_{a}$ 
value was slightly off-set from the $b$-axis as shown in 
Fig.~\ref{ExpResult} (b).  The obtained 120 K spectra show rich 
dispersing features.  In going from $k_{b}= 0$ to $\pi/2$, a single 
peak centered at $E_B=0.9$ eV is split into two features: the 
splitting becomes largest at $k_{b}= \pi/2$ with the two features 
located at $E_B\sim 0.7$ and $\sim 1.4$ eV. The 0.7 eV peak then 
decreases in intensity in going from $k_{b}= \pi/2$ to $\pi$ and only 
a single broad peak is left at $E_{B} \sim 1.1$ eV at the BZ boundary 
$k_{b}\sim\pi$.  The $k_{b}$-dependence of the spectra between $k_{b}= 
0$ and $2\pi$ is almost symmetric with respect to $k_{b} = \pi$, which 
excludes significant changes in the photoemission matrix elements 
between the first and second BZ's.  By contrast, the 300 K spectra, 
which well agree with the previous 
report~\cite{KobayashiPRL98,CommentOnPrevRep}, show less pronounced 
features than the 120 K spectra.  The spectra at $k_{b}=0$ and $\pi$ 
become a broader peak with a longer tail towards high binding 
energies.  As for the spectra at $k_{b} \sim 0.5\pi$, the peak located 
at $E_B \sim 0.7$ eV becomes weaker and that at $\sim 1.4$ eV stronger 
in going from 120 K to 300 K. We also performed ARPES measurements at 
200 K and confirmed that the changes are gradual as a function of 
temperature.  These observations are more clearly recognizable in the 
intensity plots (b) and (c) of Fig.~\ref{CompExpTheory}.

By noting that the finite-temperature effects strongly depend on the 
momentum ($k \equiv k_{b}$), they are clearly not due to simple 
thermal broadening or charging effects.  The following two points may 
be remarked: (i) in the temperature range studied here, there is no 
phase transition in this compound that may give rise to such a 
dramatic change; (ii) the energy scale of the spectral change is not 
of order $kT \sim 0.03$ eV but of order $\sim 1$ eV. These phenomena 
are obviously beyond the conventional band picture and would reflect 
strong correlation effects.

\begin{figure} [ht]
\center \epsfxsize=70mm \epsfbox{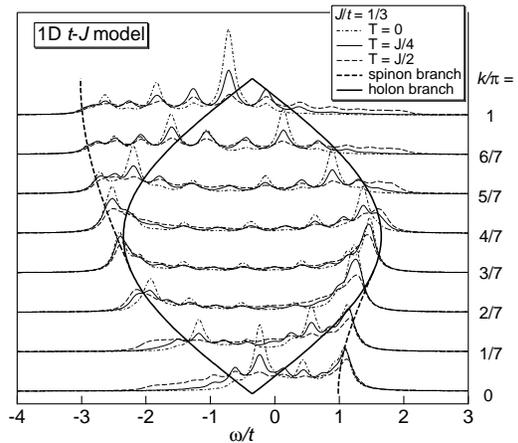} 
\vspace{1pt} 
\caption{Spectral function of the half-filled 14-site $t$-$J$ model 
with $J/t=1/3$ at various temperatures.  The thick solid and dashed 
curves indicate the singularities of the spinon and holon branches, 
respectively.}
\label{TohyamaCalc}
\end{figure}

In order to interpret the above observations quantitatively, we have 
adopted the 1D $t$-$J$ model to calculate the one-particle spectral 
function $A(k, \omega)$ at finite temperatures by the exact 
diagonalization method~\cite{TohyamaPC93}.  As the behavior of 
$\chi(T)$ well above $T_{SP}$ indicates, the $t$-$J$ model is valid as 
one of the simplest models to describe this system in the temperature 
region considered.  It may be applicable not only to the half-filled 
chain case~\cite{CarpyActaCryst77,IsobeJPSJ96-1}, but also to the 
quarter-filled ladder case where each $d$ electron is localized in a 
rung of two V atoms~\cite{SmolinskiPRL1998,OhamaPreprint}.  In the 
ladder case, only the half-filled bonding band is taken into account, 
because no electrons are in the antibonding 
band~\cite{SmolinskiPRL1998}.  The quarter-filled ladder can be mapped 
on to the half-filled chain by taking only the half-filled bonding 
band with the empty antibonding band neglected and by an appropriate 
modification of the parameters $t$ and 
$J$~\cite{SmolinskiPRL1998,HorschCM9801316}.  In addition, the $t$-$J$ 
model has the advantage over a more realistic Hubbard model in that 
the former can treat larger clusters, which is crucial for the 
discussion of finite temperature effects.  We have calculated $A(k, 
\omega)$;
\begin{eqnarray}
A(k,\omega) &=& \frac{1}{Z}\sum_{f, \sigma}e^{-\beta 
E_{i}^{N}}\vert\langle f, N-1 \vert c_{k \sigma}\vert i, N 
\rangle\vert^{2}\nonumber\\
&\times&\delta(\omega-E_{i}^{N}+E_{f}^{N-1}),
\end{eqnarray}
where $Z = \sum_{i} e^{-\beta E_{i}^{N}}$ is the partition function.  
Results for a half-filled 14-site $t$-$J$ cluster with $J/t=1/3$ at 
$T=0, J/4$ and $J/2$ are shown in 
Fig.~\ref{TohyamaCalc}~\cite{CommentOnNa}.  $A(k,\omega)$ at $T=0$ can 
be intuitively interpreted as a convolution of spinon and holon 
excitations, whose dispersional widths are $\sim J$ and $\sim 2t$, 
respectively.  In this scenario, in the ground state, the holon band 
is empty while the spinon band is half-filled up to the Fermi momentum 
$k=\pi/2$~\cite{KimPRL96}.  The lineshape of the ``spinon branch'' 
(see Fig.~\ref{TohyamaCalc}) is determined by the band edge 
singularity of the holon band whereas that of the ``holon branch'' is 
determined by the Fermi edge singularity of the spinon 
band~\cite{SuzuuraPRB97Sorella}.  At finite temperatures, the spectral 
weight of the spinon branch at $\omega/t > 1$ ($\omega/t < -2$) is 
transferred from $0<k<\pi/2$ ($\pi/2 <k<\pi$) to $\pi/2 <k<\pi$ 
($0<k<\pi/2$).  At the same time, the intensity of the holon branch 
decreases.  In fact, spectral weight is transferred from the spinon 
branch to a wide energy region, making the spectral features less 
pronounced and more symmetric with respect to $k = \pi/2$.  In fact, 
the singularity of the holon branch due to the existence of the spinon 
Fermi surface is easily smeared out over the entire energy range of 
$4t$ ($\gg T \sim J$) at finite temperatures of order 
$J$~\cite{PencPRB97}.

In Fig.~\ref{CompExpTheory} we show a comparison between the 
experimental and theoretical spectra.  Figure~\ref{CompExpTheory} (a) 
shows experimental spectra for selected momentum values and 
Figs.~\ref{CompExpTheory} (b) and (c) show the intensity plot in the 
$E$-$k$ plane on the grey scale.  Correspondingly, we show in 
Figs.~\ref{CompExpTheory} (d)-(f) the theoretical spectra at $T=J/4 
\sim 150$ K (solid lines) and $T=J/2 \sim 300$ K (dashed lines) after 
Gaussian broadening with the width $2.5t$ and their intensity plots.  
Here, we have assumed that $t= 3J = 0.15$ eV, which is plausible 
because $J = 4t^{2}/U \sim 0.05$ eV and $U = 2$-$4$ eV in typical 
vanadium oxides~\cite{UValueVanadiumOxide}.  Comparing (a)-(c) and 
(d)-(f), overall agreement is satisfactory.  In the low temperature 
region ($T=120$ K or $J/4$), the experimental shift of the peak 
position between $k_{b} =0.0$ and $0.95\pi$ may be attributed to the 
existence of the spinon branch, resulting in the asymmetry of the 
spectra with respect to $k = \pi/2$.  Besides, between $k_{b} 
=0.32\pi$ and $0.72\pi$, there are two dispersing features which may 
be assigned to the two holon branches as reproduced in the theoretical 
spectra.  These findings are also substantiated in the comparison 
between (b) and (e).  Unlike the 1D cuprates, where the intense O 
2\textit{p} structure obscures the higher binding energy part of the 
holon and spinon branches~\cite{KimPRL96}, the whole structure of 
theoretical $A(k,\omega)$ can be compared with the experimental 
results of this compound.

In the high temperature region ($T=300$ K or $J/2$), both results 
become broader and more symmetric with respect to $k=\pi/2$.  As a 
result, the tendency of the experimental spectral weight 
redistribution due to finite temperature is grossly reproduced by the 
theory as seen in the intensity plots in Fig.~\ref{CompExpTheory}.  
Around $k\sim 0$, agreement is quite excellent.  To this extent, the 
experimental finite temperature effects may be attributed to the 
existence of the spinon Fermi surface, which theoretically causes the 
dramatic spectral redistribution over the entire $E$-$k$ space with 
changing temperature.  To be more precise, however, there exist some 
discrepancies between theory and experiment.  When the temperature is 
increased from 120 K to 300 K, the spectra change more dramatically 
than the theoretical prediction.  Furthermore, while the temperature 
dependence is rather well simulated by theory at $k_{b}\sim 
0$-$0.32\pi$, at $k_{b} \sim 0.5\pi$ the feature around $E_{B}=0.7$ eV 
in experiment loses much of its spectral weight in going from 120 K to 
300 K, in disagreement with theory.  In addition, around $k_{b} \sim 
\pi$ in going from 120 K to 300 K, the experimental spectra lose 
spectral weight at $E_{B} < 0.8$ eV and a longer tail develops on the 
high binding energy side unlike the theoretical spectra.  At low 
temperatures, where the decay of a photohole is probably dominated by 
purely electronic mechanisms while other decay channels may become 
available at higher temperatures.  As a candidate, we may consider 
electron-phonon interaction and possible charge disorder, which may 
become important at higher temperatures.  The difference between the 
$t$-$J$ model and the Hubbard model as well as the degeneracy of the V 
3\textit{d} orbitals might be another origin for the discrepancy.

In conclusion, we have made an ARPES study of 
Na$_{0.96}$V$_{2}$O$_{5}$ by changing the temperature and found that a 
strong spectral weight redistribution occurs in the lower Hubbard 
band.  Also we have calculated the one-particle spectral function of 
the 1D $t$-$J$ model at finite temperatures by the exact 
diagonalization method.  The overall agreement between theory and 
experiment implies that the spin-charge separation picture is valid in 
this system.  Although they are more drastic than the theoretical 
prediction, the experimental finite temperature effects have been 
partly explained by the theory, which may be expressed as the ``Fermi 
surface'' effect of the spinon band.

We would like to thank H. Suzuura, H. Shiba, K. Penc, C. Kim, N. 
Kawakami, T. Mutou, and D. van der Marel for informative discussions.  
This work was supported by a Special Coordination Fund from the 
Science and Technology Agency of Japan.  One of us (KK) is supported 
by a Research Fellowship of the Japan Society for the Promotion of 
Science for Young Scientists.



\begin{references}
\bibitem{CommentOnPrevRep} We have obtained spectra in Fig.~3 of 
Ref.~\cite{KobayashiPRL98} by normalizing them to their peak height 
rather than the area.  It is noted that the spectra of the more 
insulating Na$_{1.00}$V$_{2}$O$_{5}$ spectra were slightly distorted 
toward higher $E_{B}$ due to weak charging effect even at 300 K.
\bibitem{CommentOnNa} Na vacancies are more sparsely distributed than 
the cluster size used for the exact diagonalization calculation, which 
justifies our quantitative comparison hereafter.
\end{references}
\end{document}